\documentclass[12pt]{cernprep}
\usepackage{graphicx}
\usepackage{amssymb}
\usepackage{epsfig,color,rotating}
\usepackage{epstopdf}
\begin{document}
\newcommand{\dedx}{\mbox{${\rm d}E/{\rm d}x$}}
\newcommand{\pT}{\mbox{$p_{\rm T}$}}
\newcommand{\GeVc}{\mbox{GeV/{\it c}}}
\newcommand{\MeVc}{\mbox{MeV/{\it c}}}
\newcommand{\nue}{\mbox{$\nu_{\rm e}$}}
\newcommand{\num}{\mbox{$\nu_{\mu}$}}
\newcommand{\anue}{\mbox{$\bar{\nu}_{\rm e}$}}
\newcommand{\anum}{\mbox{$\bar{\nu}_{\mu}$}}
\newcommand{\anuetoanum}{\mbox{$\bar{\nu}_{\rm e} \rightarrow \bar{\nu}_{\mu}$}}
\newcommand{\numtonue}{\mbox{$\nu_{\mu} \rightarrow \nu_{\rm e}$}}
\newcommand{\anumtoanue}{\mbox{$\bar{\nu}_{\mu} \rightarrow \bar{\nu}_{\rm e}$}}
\newcommand{\pip}{\mbox{$\pi^+$}}
\newcommand{\pim}{\mbox{$\pi^-$}}
\newcommand{\mup}{\mbox{$\mu^+$}}
\newcommand{\mum}{\mbox{$\mu^-$}}
\newcommand{\eplus}{\mbox{e$^+$}}
\newcommand{\eminus}{\mbox{e$^-$}}
\newcommand{\Cdouze}{\mbox{$^{12}$C}}
\newcommand{\Ndouze}{\mbox{$^{12}$N}}
\newcommand{\Bdouze}{\mbox{$^{12}$B}}
\newcommand{\Rgam}{\mbox{$R_{\gamma}$}}
\newcommand{\Rbeta}{\mbox{$R_{\beta}$}}
\newcommand{\gam}{\mbox{$\gamma$}}
\newcommand{\gams}{\mbox{$\gamma$'s}}
\newcommand{\signalreaction}{\mbox{$\bar{\nu}_{\rm e}$ + p $\rightarrow$ e$^+$ + n}}
\begin{titlepage}
\vspace*{5mm}
\begin{flushright}
15 December 2011 \\
\end{flushright} 
\vspace{1cm}
\title{\large{Reply to `Corrections to the HARP--CDP Analysis \\
 of the LSND Neutrino Oscillation Backgrounds'}}

\begin{abstract}
The alleged mistakes in recent papers that reanalyze the backgrounds to the `LSND anomaly' do not 
exist. We maintain our conclusion that the significance of the `LSND anomaly' is not 3.8~$\sigma$ but not larger than 2.3~$\sigma$.

\end{abstract}

\vfill  \normalsize
\begin{center}
The HARP--CDP group  \\  

\vspace*{2mm} 

A.~Bolshakova$^1$, 
I.~Boyko$^1$, 
G.~Chelkov$^{1a}$, 
D.~Dedovitch$^1$, 
A.~Elagin$^{1b}$, 
D.~Emelyanov$^1$,
M.~Gostkin$^1$,
A.~Guskov$^1$, 
Z.~Kroumchtein$^1$, 
Yu.~Nefedov$^1$, 
K.~Nikolaev$^1$, 
A.~Zhemchugov$^1$, 
F.~Dydak$^2$, 
J.~Wotschack$^{2*}$, 
A.~De~Min$^{3c}$,
V.~Ammosov$^{4\dagger}$, 
V.~Gapienko$^4$, 
V.~Koreshev$^4$, 
A.~Semak$^4$, 
Yu.~Sviridov$^4$, 
E.~Usenko$^{4d}$, 
V.~Zaets$^4$ 
\\
 
\vspace*{8mm} 

$^1$~{\bf Joint Institute for Nuclear Research, Dubna, Russia} \\
$^2$~{\bf CERN, Geneva, Switzerland} \\ 
$^3$~{\bf Politecnico di Milano and INFN, 
Sezione di Milano-Bicocca, Milan, Italy} \\
$^4$~{\bf Institute of High Energy Physics, Protvino, Russia} \\

\vspace*{5mm}

\end{center}

\vspace*{5mm}
\rule{0.9\textwidth}{0.2mm}

\begin{footnotesize}

$^a$~Also at the Moscow Institute of Physics and Technology, Moscow, Russia 

$^b$~Now at Texas A\&M University, College Station, USA 

$^c$~On leave of absence

$^d$~Now at Institute for Nuclear Research RAS, Moscow, Russia

$^{\dagger}$~Deceased

$^*$~Corresponding author; e-mail: joerg.wotschack@cern.ch
\end{footnotesize}

\end{titlepage}


\newpage 

In two recent papers~\cite{firstpaper,secondpaper}, we reanalyzed the backgrounds to LSND's signal of 
$\bar{\nu}_{\mu} \rightarrow  \bar{\nu}_{\rm e}$ oscillations~\cite{LSNDPRD64} (the `LSND anomaly'), questioned LSND's analysis procedures, and claimed that the significance of the `LSND anomaly' is not 3.8~$\sigma$ but 
not larger than 2.3~$\sigma$.

The concerns discussed in our papers are: 

\begin{enumerate}
\item inadequate knowledge of $\pi^\pm$ production by 800~MeV/{\it c} protons on various 
target nuclei at the time when LSND calculated their neutrino fluxes;
\item inadequate or missing consideration of pion production by higher-generation protons, pions, and---in particular---neutrons;
\item the inclusion of the first bin in the fit of the \Rgam\ distribution that does not carry any information on the distinction between correlated and accidental $\gamma$'s; 
\item questions on the `efficiencies' of correlated and accidental $\gamma$'s; 
\item questions on the effective rate of  accidental $\gamma$'s
and their influence on the \Rgam\ distributions of correlated and accidental $\gamma$'s; 
\item missing systematic errors of the `base distributions' of correlated and accidental $\gamma$'s; and
\item missing positrons from $^{12}$N$_{\rm gs}$ beta decays that are misidentified as correlated $\gamma$'s.
\end{enumerate}

Of these seven items, only items 1 and 7 are addressed in the
paper `Corrections to the HARP--CDP Analysis 
of the LSND Neutrino Oscillation Backgrounds' by G.T.~Garvey {\em et al.}~\cite{Garvey2011} that
claims several mistakes in our analysis. 

Before we argue in the following that these alleged mistakes do not exist, a clarification is in order.
LSND insist on the notion that their neutrino flux calculations are correct within their quoted errors. 
We disagree with this notion. First, the knowledge of pion production cross-sections has 
considerably improved in the
last decade (we use state-of-the-art Geant4 and FLUKA cross-sections, further improved by
pertinent experimental results that are reported in Ref.~\cite{firstpaper}). Second,
LSND apply the same percentage errors to neutrino flux integrals and to parts of the 
neutrino spectra---in particular to the small high-energy portions of the spectra, which is not realistic.
Third, the LSND parametrization of
pion cross-sections does not describe the results of their calibration experiment E866~\cite{E866}. 
All this leads to much too optimistic error assignments on neutrino fluxes by LSND, and to 
misconceptions on what quantities have actually been `measured'.

Our simulation of LSND's setup serves the following purpose:
to assess quantitatively the effect of the improvement of pion production cross-sections
over those that were employed by LSND more than a decade ago. This necessitates 
(i) a simulation that can switch between the application of the
LSND cross-sections (`LSND emulation') and our `best estimate' cross-sections; and 
(ii) a demonstration that this simulation reproduces, with LSND cross-sections,
approximately the published LSND result on the conventional $\bar{\nu}_{\rm e}$ background. 

Our `LSND emulation' reproduces 
with $0.585 \times 10^{-12}$~(pot.cm$^2$)$^{-1}$ remarkably well LSND's published result on the 
conventional $\bar{\nu}_{\rm e}$ background of
$0.65 \times 10^{-12}$~(pot.cm$^2$)$^{-1}$. Then we add the observed increase when switching from the 
`LSND emulation' cross-sections to our `best estimate' cross-sections, to the conventional $\bar{\nu}_{\rm e}$ background that LSND published. In this procedure small differences 
between the real LSND setup
and its simulation in our program cancel,  and we obtain the 
conventional $\bar{\nu}_{\rm e}$ background that LSND 
should have obtained with our `best estimate' cross-sections. The stringent result is
an increase by a factor of 1.6.

The $\bar{\nu}_{\rm e} /  \bar{\nu}_\mu$ ratio is not a useful quantity for the normalization of neutrino fluxes 
as $\bar{\nu}_{\rm e}$ originate from $\mu^-$ from $\pi^-$ decays while $\bar{\nu}_\mu$ originate 
from $\mu^+$ from $\pi^+$ decays.
In the LSND energy domain $\pi^+$ production is essentially unrelated with $\pi^-$ production. 
Therefore, it makes no sense to introduce an unnecessary error
by transporting the uncertainty in $\pi^+$ production into the calculation of the conventional
$\bar{\nu}_{\rm e}$ background which originates from $\pi^-$ only. 

Our `best estimate' of $\nu_\mu$ and $\bar{\nu}_\mu$ rates from pion decay in flight 
is larger than the LSND estimate by factors of 2.5 and 3.3, respectively. 
The $\nu_\mu$ and $\bar{\nu}_\mu$ rates refer to $E_\nu > 123.7$~MeV for
$\nu_\mu$ and to $E_\nu > 113.1$~MeV for $\bar{\nu}_\mu$ and hence to very small high-energy portions
of the respective neutrino spectra.
These portions are not relevant for the calculation of the 
conventional $\bar{\nu}_{\rm e}$ background. Rather, they are relevant for 
(i) calculating what we termed `Background II' that concerns primarily events where a muon is misidentified as electron, and 
(ii) to appreciating LSND's constraints on neutrino fluxes (which in no way
contradict our claim of a conventional $\bar{\nu}_{\rm e}$ rate larger by a factor of 1.6). 
The factor of 3.3 for $\bar{\nu}_\mu$ is essentially unrelated with our claim of a 
$\bar{\nu}_{\rm e}$ background larger by a factor of 1.6 than calculated by LSND. The
proportionality between these two quantities purported by Garvey {\em et al.\/}~\cite{Garvey2011}
lacks justification.

We agree that `3 MeV' is not a hard cutoff for muons that are not observed. Rather, it is a reasonable estimate. 
It is for this reason that `3 MeV' is nowhere used in our calculation of `Background II'. We consider it safer to calculate Background II from the small portion of the neutrino spectrum between the neutrino energy threshold and 4~MeV above. The value of
4~MeV is approximate and beset with instrumental uncertainties; we bypass these 
uncertainties by considering not
absolute numbers but relative numbers: we determine a factor by dividing `best estimate'
predictions by `LSND emulation' predictions, and apply this factor as a correction to
LSND's estimate of Background II.

The final LSND physics paper~\cite{LSNDPRD64} comprises a discussion 
of backgrounds that mimic 2.2 MeV $\gamma$'s. There, even backgrounds as low as 
0.1 events are discussed. There is no mention of a background from $^{12}${\rm N}$_{\rm gs}$
beta decays that is now claimed to be $\sim$0.2 events.
As for the claim that an electron produces many more PMT hits than a photon, we note that
the energy of a 2.2 MeV photon is in mineral oil within a few cm fully converted into the
kinetic energy of electrons by Compton scattering and photoelectric interaction.
The $R$ distribution of betas from $^{12}${\rm N}$_{\rm gs}$ decays 
can neither agree with the \Rgam\ distribution of
correlated $\gamma$'s nor with the \Rgam\ distribution of uncorrelated $\gamma$'s, for the
pulseheight distribution of the beta's is intrinsically different. It is straightforward
to calculate the background from $^{12}${\rm N}$_{\rm gs}$ beta decays that mimic correlated $\gamma$'s, and we stand by our estimate of a background of 2.3 events that was neglected by LSND. 

In summary, we maintain our conclusion that the significance of the `LSND anomaly' is not 3.8~$\sigma$ but 
not larger than 2.3~$\sigma$.

\end{document}